\documentclass[11pt]{article}

\usepackage{epsfig, url, amsmath, amssymb, bm, accents, fullpage}
\usepackage{fullpage,setspace, natbib, verbatim, setspace, rotating,
verbatim}
\usepackage{subfigure}
\usepackage{float}
\usepackage{color}
\usepackage{fancybox}
\usepackage{rotating}
\usepackage{hyperref,amsthm}
\def\argmax{\operatornamewithlimits{arg\,max}}
\usepackage[margin=1.25in]{geometry}

\newtheorem*{remark}{Remark}

\DeclareMathOperator{\tr}{tr}
\DeclareMathOperator{\diag}{diag}

\DeclareMathOperator{\iid}{\stackrel{\mbox{\tiny iid} }{\sim}}
\DeclareMathOperator{\ind}{\stackrel{\mbox{\tiny ind} }{\sim}}

\begin{document}

\title{A symmetric prior for multinomial probit models}

\author{
Lane F.\ Burgette\thanks{RAND Corporation}
\and 
David Puelz\thanks{The University of Chicago, Booth School of Business}
\and
P.\ Richard Hahn\thanks{Arizona State University}
}

\date{}
\maketitle
\vspace{5pt}

\begin{center}
\textsc{This Version: \today}
\end{center}

\thispagestyle{empty}

\singlespacing
\begin{abstract}
Fitted probabilities from widely used Bayesian
multinomial probit models can depend strongly on the choice of
a base category, which is used to uniquely identify the parameters of the model.  
This paper proposes a novel identification strategy, and associated prior
distribution for the model 
parameters, that renders the prior symmetric with respect to relabeling
the outcome categories.  The new prior permits an efficient Gibbs algorithm that
samples rank-deficient covariance matrices without resorting to
Metropolis-Hastings updates.
\end{abstract}


\setcounter{page}{0}\thispagestyle{empty}\baselineskip18.99pt\newpage 


\newpage
\doublespacing


\section{Introduction}
In multinomial probit (MNP) models of discrete choices, parameters are
typically identified by selecting a base category relative to which
the choice parameters are defined. From the point of view of
identification, the choice of base category is immaterial.  However,
in a Bayesian framework,  base category specification affects the prior predictive choice probabilities, which in turn affects posterior inference --- sometimes strongly so.
 
In this paper, we propose sum-to-zero restrictions on the latent
utilities and regression parameters that define the MNP model. Under this
novel identification 
framework, we are able to develop a prior which is symmetric with respect to
relabeling of the outcome categories.  We show that this new parametrization and the associated prior preserve
the favorable computational aspects of other, recent Bayesian MNP
models \citep{ivd, burgetteNord, jiao2015corrected}. 


\subsection{Multinomial probit models of discrete choice}
Multinomial probit (MNP) models are popular in studies involving 
discrete choice data \citep{mcfadden74, train}.  They have applications in
marketing \citep{rossiBook}, politics \citep{rudolph}, 
transportation studies \citep{mcfadden74, garrido}, and beyond.
The MNP is  more flexible than 
standard multinomial logit models, as it
need not make an assumption of independence of irrelevant
alternatives (IIA).  This means that the ratio of selection
probabilities for two outcome categories can depend on the characteristics
of another category.
Further contributing to the popularity of the MNP is a series of 
advances in Bayesian computation, starting with \cite{albertChib},
that has made it increasingly computationally manageable \citep{mcRossi,
  mcEtAl, ivd, ivdComp}.

The MNP requires two normalizations in order to
identify the model.  These models can be derived through the
assumption that agents construct latent Gaussian utilities and select
the category that corresponds to the largest utility.  Since the
ordering of the utilities is maintained by an additive shift or
multiplicative rescaling, 
identifying assumptions on the scale and location are
needed.  

In order to set the scale, it has been standard to fix an
element on the main diagonal of the covariance matrix at one.  
\cite{burgetteNord}  demonstrated that the choice of which element
one fixed 
could have a meaningful impact on posterior predictions, when using the
popular prior of \cite{ivd}.  To avoid this
problem, they proposed a model that identifies the scale of the model
by fixing the trace of the covariance matrix, which makes the prior
covariance  invariant to
joint permutations of the rows and columns.  
This paper will build upon such a trace-restricted prior, resolving
the location identification issue as well.

Previous MNP models have set the location of the
latent utilities by specifying a base (or reference) category for
the model. The base category's utility is then subtracted from all of
the other utilities for each observation,
removing the indeterminacy of the location.   
But, \cite{burgetteNord} noted that Bayesian MNP predictions can
be sensitive to
the specification of the base category, though they did not provide a
satisfactory solution for this issue.  This problem arises because
instead of specifying a prior for the original utilities and inducing
a prior on the base-subtracted utilities, it has been standard to
specify a prior directly on base-subtracted utilities.

Rather than selecting a reference
category whose utility is assumed to be equal to zero, we enforce a
sum-to-zero restriction on the latent utilities.  
If respondents choose from $p$ categories, other MNP methods
transform the utilities to $(p-1)$-space.  Instead, we constrain our
utilities to exist in a $(p-1)$-dimensional hyperplane in $p$-space.

We apply our new prior
 to two  consumer choice datasets, as well as a series of
simulated datasets based on the consumer choice studies.  In doing so,
we see that the {\em symmetric MNP} (sMNP) model
defines a more sensible model, produces better predictions, and has
favorable computational properties compared to previous MNP models.  


\subsection{Preliminaries}

Assume that agent $i = 1, \ldots, n$ is choosing
among $p$ mutually exclusive alternatives.  
The MNP can be derived by assuming that there exist vectors of 
latent Gaussian utilities $W_i = \{w_{ij}\}$ of length $p$, 
and that each agent selects the alternative with
the highest utility, so that we observe $Y_i = \arg\max_j w_{ij}$.  

It is standard to assume that the utilities take the form 
\begin{equation}
W_{i} = X_i\beta + \varepsilon_i.
\end{equation}
$X_i$ is a matrix of covariates, $\beta$ is a vector of regression
parameters, and $\varepsilon_i\iid \mbox{normal}(0,\Sigma)$ capture
variations in taste across agents.  
We will assume $X_i$ contains intercept terms, $k_d$
covariates that vary by decision-maker (e.g., a buyer's age),
and $k_a$  
alternative-specific covariates (e.g., product prices).  
We assume the covariates are arranged in that order (from left to
right) so that  
\begin{equation}
X_i = \begin{bmatrix}I_p & (x_{i}^d)^\top \otimes I_p & x_{i}^a \end{bmatrix}.
\end{equation}
The $k_d$-vector $x_{i}^d$ is the collection of covariates that vary
by individual; $x_{i}^a$ is a $p\times k_a$ 
matrix whose columns contain the values
of the variables that vary by alternative.  In more detail,
$$w_{ij} = \eta_j + (x_{i}^d)^\top\xi_j + x_{ij}^a\delta + \epsilon_{ij},$$ so that $\beta^T = (\eta_1, \dots \eta_p, \xi_1^T, \dots \xi_p^T, \delta^T)$, making $\beta$ a length $p + (p \times k_d) + k_a$ vector. 

A standard identifying approach (cf. \cite{rossiBook}, section 4.2) is to transform $W_i$ to
$W_i^* = T_{\mbox{\tiny bc}}W_i$ where  
\begin{equation}
T_{\mbox{\tiny bc}} = \begin{bmatrix} -J_{p-1} & I_{p-1} \end{bmatrix}
\label{eqn:Tbc}
\end{equation}
with $J_{p-1}$ a column vector of ones with length $p-1$.  This amounts to choosing the first category as the base category (without loss of generality) and subtracting it from the other utilities. For $j > 1$, this gives
\begin{equation}
\begin{split}
w_{ij}^*  &=   w_{ij} - w_{i1},\\
  &=   \eta_j + (x_{i}^d)^\top\xi_j + x_{ij}^a\delta + \epsilon_{ij} - (\eta_1 + (x_{i}^d)^\top \xi_1 + x_{i}^a\delta + \epsilon_{i1})\\
  &= \eta_j - \eta_1 + (x_{i}^d)^\top (\xi_j - \xi_1) + (x_{ij} - x_{i1})\delta + (\epsilon_{ij} - \epsilon_{i1}).
\end{split}
\end{equation}
It follows that $W^*_i = X_i^*\beta^* + \varepsilon_i^*$
where
\begin{equation}
X^*_i = \begin{bmatrix}I_{p-1} &  (x_{i}^d)^\top\otimes I_{p-1} & T_{\mbox{\tiny bc}}x_{i,a} \end{bmatrix},
\end{equation}
\begin{equation}
\beta^* = (\eta_2 - \eta_1, \dots, \eta_p - \eta_1, (\xi_2 - \xi_1)^\top \dots (\xi_p - \xi_1)^\top, \delta),
\end{equation}
and $\varepsilon_i^* \iid \mbox{normal}(0, \Sigma^*= T_{\mbox{\tiny bc}}\Sigma T_{\mbox{\tiny bc}}^\top)$.  Under this parametrization, $Y_i = \arg\max_j w^*_{ij} + 1$ if $w^*_{ij} >0$ and $Y_i = 1$ if $\max_j w^*_{ij} < 0$.  

\cite{albertChib} had the key insight that data augmentation  
\citep{tannerWong} would greatly ease the estimation of the MNP.  
If we treat the latent $W_i^*$ as parameters to be updated in the MCMC
algorithm, then under a normal prior, 
the full conditional distribution of $\beta^*$ is
normal.
Further, the full conditional distribution of each $W^*_i$ is
truncated multivariate normal, which can be updated one component at a
time as univariate truncated normals \citep{mcRossi}.  

It then remains to sample $\Sigma^*$, the ($p-1$)-dimensional covariance over the base-subtracted utilities.
Up to a constraint and the normalizing constant, 
the priors for both the Imai and van Dyk and the Burgette and Nordheim
models are the same:
\begin{equation}
p(\Sigma^*)\propto |\Sigma^*|^{-(\nu +
  p)/2}[\tr(S{\Sigma^*}^{-1})]^{-\nu(p-1)/2}\bm 1\{\mbox{cond}\},
\label{bnPrior}
\end{equation}
where $\bm 1\{\mbox{cond}\}$ is equal to one if $\{\mbox{cond}\}$ is a
true statement, and 
zero otherwise.  For Imai and van Dyk, this condition is
$\{\sigma^*_{11}=1\}$; 
for Burgette and Nordheim the condition is $\{\tr(\Sigma^*) = (p-1)\}$.
Further, \cite{burgetteNord} introduce the so-called working parameter, $\alpha$, defining an unconstrained covariance $\tilde\Sigma = \alpha^2\Sigma^*$. The parameter pair $(\alpha, \Sigma^*)$ is given a joint prior
\begin{equation}
p(\Sigma^*, \alpha^2) \propto |\Sigma^*|^{-(\nu + p)/2}
\exp\{-1/(2\alpha^2) \tr(S{\Sigma^*}^{-1})\} (\alpha^2)^{-[\nu(p-1)/2
+ 1]} \bm 1\{\mbox{cond}\},
\end{equation}
where $S$ is a prior parameter, and under which posterior draws of $\Sigma^{*}$ can be obtain via a Gibbs sample 
of $\tilde\Sigma$.  

\cite{fong2016bayesian} handle this identification problem by restricting the covariance to a correlation matrix.  In their sampler, they use a Metropolis-Hastings step to first generate a covariance and then accept the implied correlation matrix with a specified acceptance probability. However, as in previously mentioned approaches, the base category is must be chosen first, and this choice can impact materially posterior inferences.  The focus of this paper is to document prior asymmetries that result from the choice of base category and to propose a new model that does not require that such a choice be made.

\subsection{Asymmetries of commonly-used MNP priors}

\begin{figure}[H]
\centering
\includegraphics[scale=0.3]{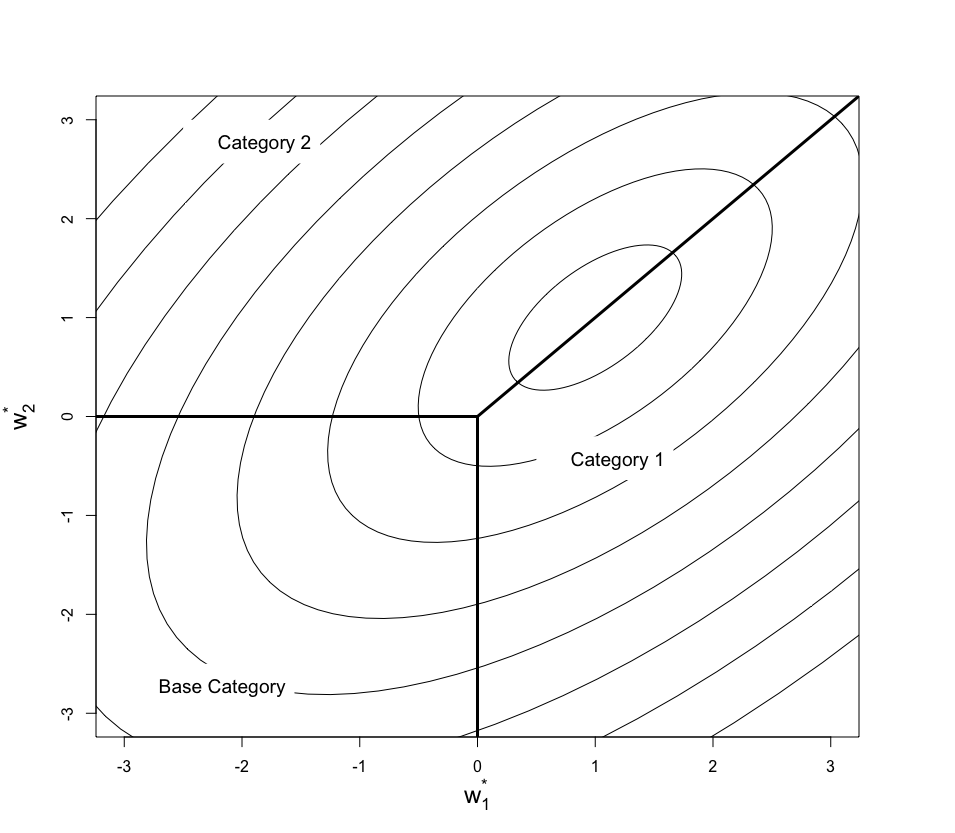}
\singlespacing{\caption{A depiction of the multivariate normal contours associated to the base-subtracted utility space for $p -1 = 2$. The base category standardization entails that the area in utility space allocated to the base category is a different shape than the regions allocated to non-base categories, meaning that standard priors over $\Sigma^*$ (which govern the contours) will result in asymmetric priors on the implied choice probabilities, which correspond to the probability, according to the prescribed multivariate normal distribution, of being in the various sectors associated with each category.}}\label{fig:contours}
\end{figure}

Later in this paper, we will demonstrate empirically that switching
from one base category to another can result in substantial
differences in {\em posterior} purchase probabilities in
marketing applications that appear elsewhere in the literature.  In this section, we highlight how such differences arise in the prior purchase probabilities under different base category specifications,
conditional on a range of values of the 
structural portion of the utilities, $X^*_i\beta^*$.  The base category standardization imposes an inherently asymmetric mapping from the utility space to probabilities, as depicted in Figure (\ref{fig:contours}). As such,  standard priors on $\Sigma^*$ will generally correspond to asymmetric distributions over choice probabilities, which we demonstrate now.

Consider a simple case with $p = 3$ categories and focus on one of
the three outcome categories, which we will refer to as the ``category
of interest'' (which is fixed).  First, we consider a specification where the category of interest
is the base category (denoted by $Y_i = 0$). Then, we consider a specification where 
the category of interest is the first non-base category (denoted by
 $Y_i = 1$). 
 
 Our experience
indicates that sensitivity to the base category primarily comes from
the prior on $\Sigma^*$ (rather than $\beta^*$), so we will condition on
$\beta^*$ in order to clarify the issue. Specifically, consider $X_i \beta = (v, 0, 0)$ for an arbitrary value of $v$ and let the first category be the category of interest. Then, if category 1 is the base category, we have $X_i^* \beta^* = (-v, -v)$ corresponding to categories 2 and 3; and when category 2 is the base category we have $X_i^*\beta^* = (v, 0)$ corresponding to categories 1 and 3.

\begin{figure}[H]
\centering
\subfigure{\includegraphics[scale=0.7]{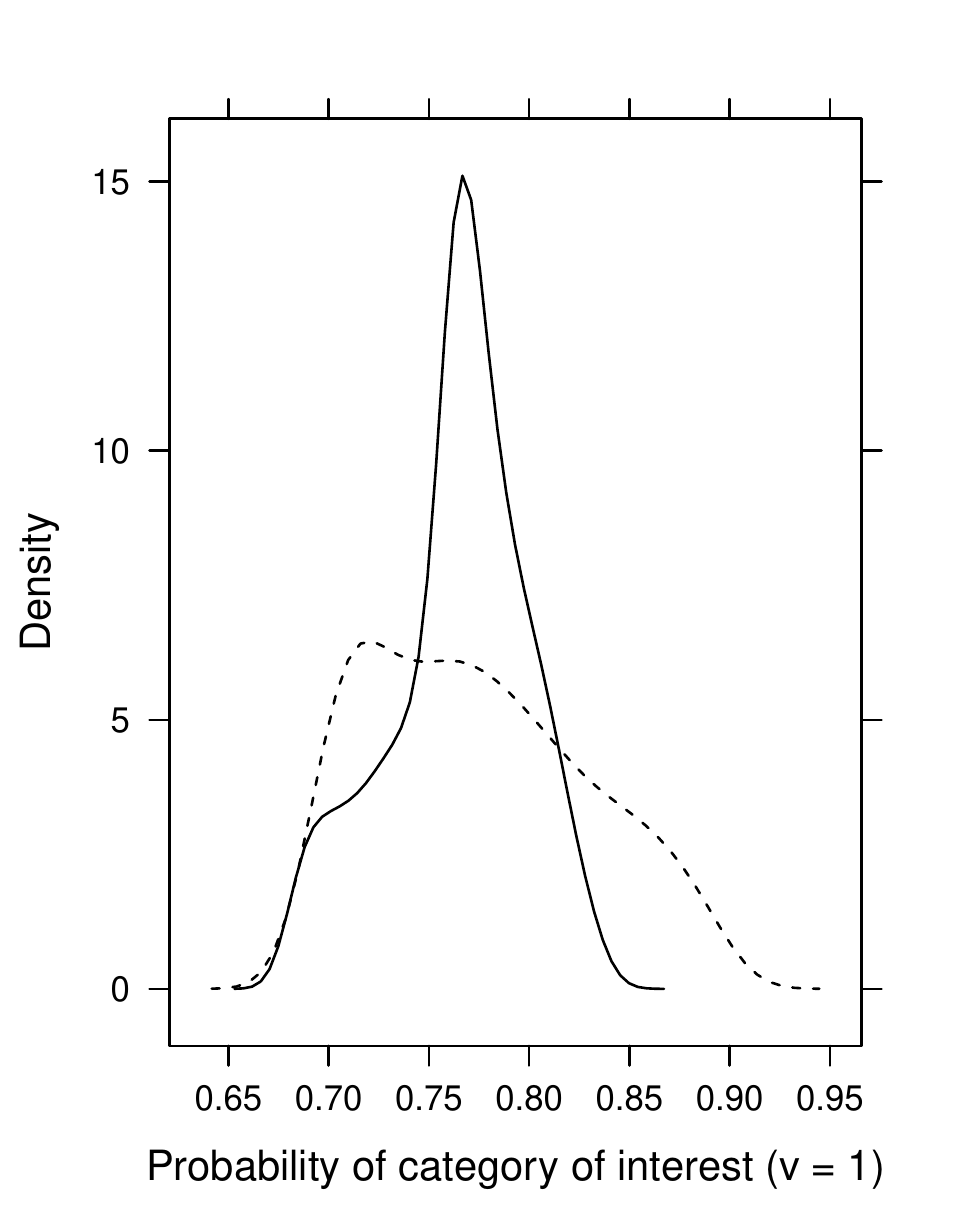}}\quad
\subfigure{\includegraphics[scale=0.7]{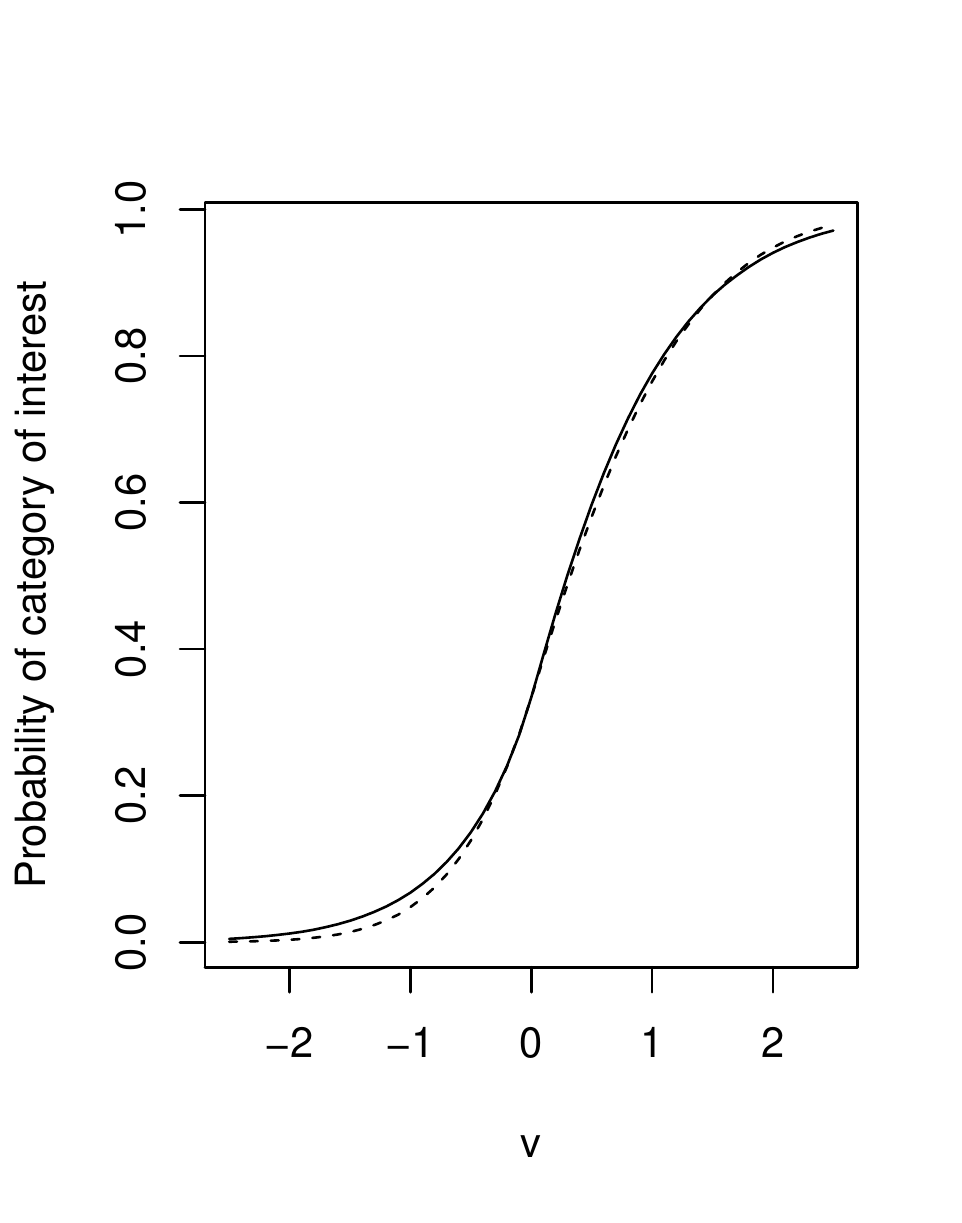}}
\singlespacing{\caption{The left-hand panel displays prior densities of the
  probability for the ``outcome of interest'' when it is coded as the
  base category (solid) and not the base category (dashed) for a
  particular value of $\beta$.  More precisely, it is the prior density of 
  $\varphi_j(1;\Sigma^*)$ for $j = 0,1$, where $j=0$
  corresponds to the solid curve, and $j=1$ corresponds to the dashed
  curve.  See expressions (\ref{varphi0}) -- (\ref{varphiEnd}) in the main text.  The
  right-hand panel plots $\psi_j(v)$ 
  across a range of $v$. That is to say, the average values of the
  distributions in the left-hand panel
  correspond to the values in the right-hand panel at $v=1$.}\label{fig:intro}}
\end{figure}

Our interest is in the quantities
\begin{eqnarray}
\varphi_0(v; \Sigma^*) &=& \Pr (Y_i = 0 \mid X_i^*\beta^* =
(-v,-v)^\top, \Sigma^*)\label{varphi0}\\ 
\varphi_1(v; \Sigma^*) &=& \Pr(Y_i = 1 \mid X_i^*\beta^* = (v,0)^\top,
\Sigma^*)\label{varphi1} \\
\psi_j(v) &=& \int \varphi_j(v; \Sigma^*)p(\Sigma^*)d \Sigma^* \quad\mbox{
 for } j=0,1. \label{varphiEnd}
\end{eqnarray}
Note that (\ref{varphi0}) and (\ref{varphi1}) both denote the probability that the category of interest is selected, but under different specification of the base category. In (\ref{varphiEnd}), $p(\Sigma^*)$ refers to the trace-restricted
variant of the Imai and van Dyk prior 
for $\Sigma^*$ with $\nu =2$ degrees of freedom, and centered at 
$S = .5J_2J_2^\top + .5I_2 \propto T_{bc} T_{bc}^\top$, with ones on the diagonal.  

Figure \ref{fig:intro} compares $\varphi_j$ and $\psi_j$ for
$j=0,1$. From the left-hand panel, note that there are strong
differences in the range of probabilities for the outcome of interest
that are supported by the prior for $\Sigma^*$, after conditioning on
$\beta^*$.  In particular, the distribution 
probabilities for the base category (solid curve) has a very sharp mode, and is
less diffuse in general relative to distribution for the nonbase category (dotted
curve).  On the other hand, the curves in the
right-hand figure nearly coincide with one another.  This indicates
that differences between the two parametrizations in the prior are
obscured by marginalizing over the distribution of $\Sigma^*$.
However, we note that these curves often times do {\em not} coincide
after conditioning  
on observed data, as will be shown in Figures \ref{fig:comp} and
\ref{fig:housePriceProb}.

Because the differences in probabilities
appear primarily to be of second and higher moments, an ad-hoc solution to the
problem of base category dependence (such
as specifying alternative values of the hyperparameters, or by specifying a
different $p(\beta^* | \Sigma^*)$ to compensate) may be difficult.
Although we expect the impact of the prior to fade
as the sample size increases, information in multinomial models
accrues slowly relative to standard models of a continuous outcome,
which means that asymmetries in the prior for an MNP model may persist
in the posterior for sample sizes that are typical in business and
economics applications.  Hence, we pursue a prior that is identically invariant to
relabeling the outcome categories.

\section{A symmetric prior for MNP regressions}
We now propose a {\em symmetric MNP} (sMNP) model that is 
invariant 
under relabeling or reordering of the outcome categories.
Rather than identifying the locations of the latent utilities by
subtracting one from the others, we instead require that they sum to
zero. (This assumes that the choice-specific covariates have mean zero
for each observation, which is a convenient but inessential
standardization.)   Further, we assume that the regression parameters
that correspond to each agent-specific covariate sum to zero, which
gives the same degrees of freedom as the standard MNP, where (in a
sense) the regression parameters related to the base category are set
equal to zero.  

With this sum-to-zero restriction on the utilities, we require a
covariance  for $W_i$ that is symmetric and positive-semidefinite
with $p-1$ positive eigenvalues,
and constrained in some way in order to set the scale of the model.
Rather than directly specifying a distribution on $p\times p$ matrices, we
build it up with a mixture of trace-restricted positive-definite
matrices.  Conditionally, we assume that a positive-definite matrix of
dimension $p-1$ describes the covariance of all but one of the
dimensions of $W_i$.  We denote the left-out category with the
parameter $b$, and refer to it as the {\em faux base category}
indicator.  In contrast to previous MNP models, $b$ 
is learned
according to Bayes rule.  

The proposed model is as follows:
\begin{eqnarray}
b & \sim & \mbox{unif}(\{1,\ldots, p\})\\ 
\Sigma_{b} &\sim& p_{\mbox{\tiny TR}}( S_b, \nu_b)\\
R_{b} & = & [\mbox{chol}(\Sigma_{b})]^\top\\
R & = & \begin{bmatrix}
R_{1:(b-1)}\\
R_b^*\\
R_{b:p}
\end{bmatrix}\\
\beta_{b} &\sim& \mbox{normal}(0,A)\\
\beta &=& f(\beta_{b}) \\ 
W_i & \ind & \mbox{normal}(X_i\beta,  RR^\top)\\
Y_i & = & \argmax_j W_i.
\end{eqnarray}

Here, $p_{\mbox{\tiny TR}}$ refers to the trace-restricted variant of the
\cite{ivd} prior in (\ref{bnPrior}) with $\{\tr(\Sigma_b) = (p-1)\}$.  Its
hyperparameters $S_b$ and $\nu_b$ may change with $b$
but we recommend using common
hyperparameters in most cases, since $S_b= 
\diag\{(1+c,\ldots,1+c)\} - c J_{p-1}J_{p-1}^\top$ for all $b$ and a common $\nu_b$ 
will yield a prior covariance structure that
is symmetric with respect to the outcome categories. \cite{burgetteNord} discuss tradeoffs for different hyperparameter choices in the trace-restricted prior.  Following their guidance, we choose a default value of $\nu_b = p+1$.  This choice provides sufficient regularization without being too informative. This corresponds to the first $p-1$ rows and columns of a symmetric
$p\times p$ covariance matrix $P$ with $p -1$ positive eigenvectors
that is symmetric with respect to relabeling the rows and columns.
This matrix has the property that vectors drawn from the
$\mbox{normal}(0, P)$ distribution sum to zero almost surely, which
is a natural center for our relabeling invariant, sum-to-zero MNP.  Using
$c=0$ means roughly that we expect $p-1$ of the dimensions of the
utilities to be independent, with the remaining dimension strongly
anti-correlated.  We recommend using $c=1/(p-1)$ since it is a more neutral prior and seems
to lead to better mixing in the MCMC. 

$R_{b}$ is the transposed Cholesky decomposition of $\Sigma_{b}$ such that
$R_{b}R_{b}^\top = \Sigma_{b}$.  $R_b^*$ is a row vector inserted
into $R_{b}$ at the 
$b$th row such
that the sum of each column of $R$ is zero.  In this formulation,
$\beta_{b}$ has dimension $(p-1)(k_d+1) + k_a$ (assuming that intercept
terms are included in the matrix of covariates as stated in Section 1.2). The function $f$ acts
on $\beta_{b}$ such that for each sub-vector of length $p-1$ that
corresponds to an
agent-specific covariate (or the intercepts), $\beta$ is equal to $\beta_{b}$
with an extra
dimension inserted at the $b$th position in the sub-vector.  This
inserted element is chosen so that the sub-vector 
sums to zero. With this model specification, we induce a prior distribution on the set of
positive-semidefinite  matrices of dimension $p$ that have exactly
$p-1$ positive eigenvalues.\footnote{It would also be possible to work with a
matrix decomposition like $\Sigma = A D A'$, where $A$ is a $p\times
(p-1)$ orthogonal matrix and $D$ is diagonal. One could then
define a prior on the Stiefel manifold that contains $A$ \citep{hoff}. 
This would be a more direct definition on positive semidefinite 
matrices, but 
inducing a prior in the manner implied by our model is conceptually
simple and guarantees favorable computational properties.}  

To make the motivation of this new set of identifying restrictions
explicit, we note that they result from transforming the unnormalized
utilities not by $T_{\mbox{\tiny bc}}$ as in (\ref{eqn:Tbc}), but rather
multiplying them by a $p$-dimensional square matrix $T_s$ that is
defined to have 
ones on the 
main diagonal, and entries of $-1/(p-1)$ elsewhere.  Note that
$\arg\max W_i = \arg\max T_s W_i$, while the elements of 
$T_s W_i$ sum to zero.  This transformation also induces
the proposed identifying restrictions on $\beta$.  If we partition
$\beta = (\beta_d, \beta_a)$, where $\beta_a$ corresponds to the
covariates that vary by outcome category,
we have 
\begin{equation}
T_s X_i \beta = X_i 
\begin{bmatrix}
(I \otimes T_s)\beta_d\\
\beta_a
\end{bmatrix}.
\end{equation}
This transformed version of $\beta$ (i.e., the second factor on the
right-hand side of the above equation) conforms to the proposed identifying
restrictions.  Similarly, a normal distribution with mean zero and
covariance $T_s\Sigma T^\top_s$ results in draws that sum to zero
almost surely.  (Note that $T_s$ is almost idempotent in the sense
that $T_sT_s = cT_s$ for some scalar $c$.  The first $p-1$ rows and
columns of $T_s$ therefore serve as our default for $S_b$ since this
corresponds to the transformed variance of $\varepsilon_i$ if its
variance in the unnormalized scale is proportional to the identity.) 


We emphasize that there is nothing inherently wrong with using the
asymmetric identifying transformation $T_{\mbox{\tiny bc}}$.  If we do
not wish for our inferences to depend on the base category, however,
the prior must compensate for the asymmetries in the transformation.
This seems quite difficult to achieve, especially if we hope to have a
computationally tractable model.  Using $T_s$, however, we can
decouple prior specification and model identification, all while
preserving the favorable computational characteristics of existing MNP
models.

\subsection{Model estimation}

We propose a Gibbs sampler to estimate the model by constructing a Markov chain on a transformed space: $(\alpha, \Sigma_b, b, W, \beta_b) \mapsto (\alpha, \Sigma_b, b, \tilde{W} = \alpha W, \tilde{\beta}_b = \alpha \beta_b)$. By explicitly working in the $(\alpha, \Sigma_b, b, \tilde{W}, \tilde{\beta}_b)$ parametrization in specifying the Gibbs sampler we avoid the mistakes discussed in \cite{jiao2015corrected}, although our algorithm is different than theirs. 

\begin{remark} Note that at every iteration in the Markov chain, $\Sigma_b$ is restricted to satisfy the identifying trace restriction. \end{remark}

\begin{remark} For brevity, the notation $\tilde{\Sigma}_b$ and $\tilde{\beta}_b$ (respectively $\Sigma_b$ and $\beta_b$) obscures the fact that there are in fact $p$ entities in our parameter space, one for each possible value of $b = 1, \dots, p$. However, given $b$ (the ``working base category''), this collection of parameters only appears in the likelihood via $\Sigma = g(\Sigma_1, \dots, \Sigma_p, b) = RR^T$ as defined in (14) and (15) and $\beta = f(\beta_1, \dots, \beta_p, b) = f(\beta_b)$ as in (17). As such, it does little harm to consider only the element of $(\tilde{\Sigma}_1, \dots, \tilde{\Sigma}_p)$ and $(\tilde{\beta}_1, \dots, \tilde{\beta}_p)$, respectively, $(\Sigma_1, \dots, \Sigma_p)$ and $(\beta_1, \dots, \beta_p)$, corresponding to the current value of $b$ in the Markov chain. \end{remark}

Let
$X_{i,b}$ indicates $X_i$ with the $b$th row and the columns specific
to the $b$th category removed.
We initialize the latent utilities $W_i$ by sampling a standard
normally-distributed vector of length $p$ and centering it at zero. 
We then permute its elements so that the maximum of each
$W_i$ coincides
with the observed $Y_i$.  

The sampler proceeds in three steps:

\begin{enumerate}
\item Draw $\tilde{W} \mid Y, \tilde{\beta}_b, b, \Sigma_b, \alpha$.
\item Draw $\tilde{\beta}_b \mid Y, b, \Sigma_b, \tilde{W}, \alpha$.
\item Draw $\alpha, \Sigma_b, b \mid Y, \tilde{\beta}_b, \tilde{W}$.
\end{enumerate}

Note that all variables referenced in these Gibbs steps are from the {\em same} parameterization: $(\alpha, \Sigma_b, b, \tilde{W}, \tilde{\beta}_b)$.  We give detailed expressions for each conditional distribution in the appendix.  For the draw of the latent utilities $\tilde{W}$, we note that in the original parametrization we have $$p(W, \alpha \mid \Sigma_b, \beta_b, Y, b) \propto I(W, Y)p(W; \beta_b, \Sigma_b)p_{TR}(\alpha \mid \Sigma_b, b).$$ By definition, $\tilde{W} = \alpha W$ and $I(W, Y)$ is the indicator that the utilities and data match correctly, so with $\alpha$ known at this stage of the Gibbs sampler, we write the conditional distribution as:
$$p(\tilde{W} \mid \Sigma_b, \tilde{\beta}_b, Y, b,\alpha) \propto I(\tilde{W}/\alpha, Y)p(\tilde{W}; \tilde{\beta}_b, \alpha^2\Sigma_b).$$ To sample $\tilde{W}$,
we iterate one-by-one through the elements of $\tilde{W}_{i,b}$. Note that $\tilde{w}_{i,b}$ is known given $b$ and $\tilde{W}_{i,b}$.
  After dropping the $b$th element of $\tilde{W}_i$ and the
  corresponding elements in $X_i$ and $\beta$, the full conditionals
  of elements of $\tilde{W}_{i,b}$ are truncated univariate normal.  The
  conditional means and variances can be calculated as described by
  \cite{mcRossi}, using $\tilde{\beta}_b$ as the coefficient vector and $\alpha^2\Sigma_b$ as the covariance. These truncations are given in the appendix. 
  
  The second draw of the transformed coefficients $\tilde{\beta}_b$ is from a normal distribution whose conditional mean and variance are provided in the appendix. The third step is a draw of the triplet of parameters $(\alpha, \Sigma_b, b)$.  This is divided into a multinomial draw for $b$ (with the variance integrated out) followed by a draw of an intermediate quantity for the variance, $\tilde\Sigma_{b}$.  Finally, $\alpha$ and $\Sigma_{b}$ are obtained by first setting $\alpha = \sqrt{\tr(\tilde\Sigma_{b})/(p-1)}$, followed by $\Sigma_b = \tilde{\Sigma}_{b}/\alpha^2$. In this way, $\tr(\Sigma_b) = p - 1$ at each iteration as in \cite{burgetteNord}.  
  
  Having obtained samples from the Markov chain defined over $(\alpha, \Sigma_b, b, \tilde{W}, \tilde{\beta}_b)$, one can transform, per-iteration,  back to the original space to obtain samples of $W = \tilde{W}/\alpha$ and $\beta_b = \tilde{\beta}_b/\alpha$.  In the following section, we demonstrate this methodology on two consumer choice data sets as well as investigate its properties with a simulation study.

\section{Demonstrations}

\subsection{Clothes detergent purchases}

\begin{figure}[H] 
\centerline{\includegraphics[scale=0.7]{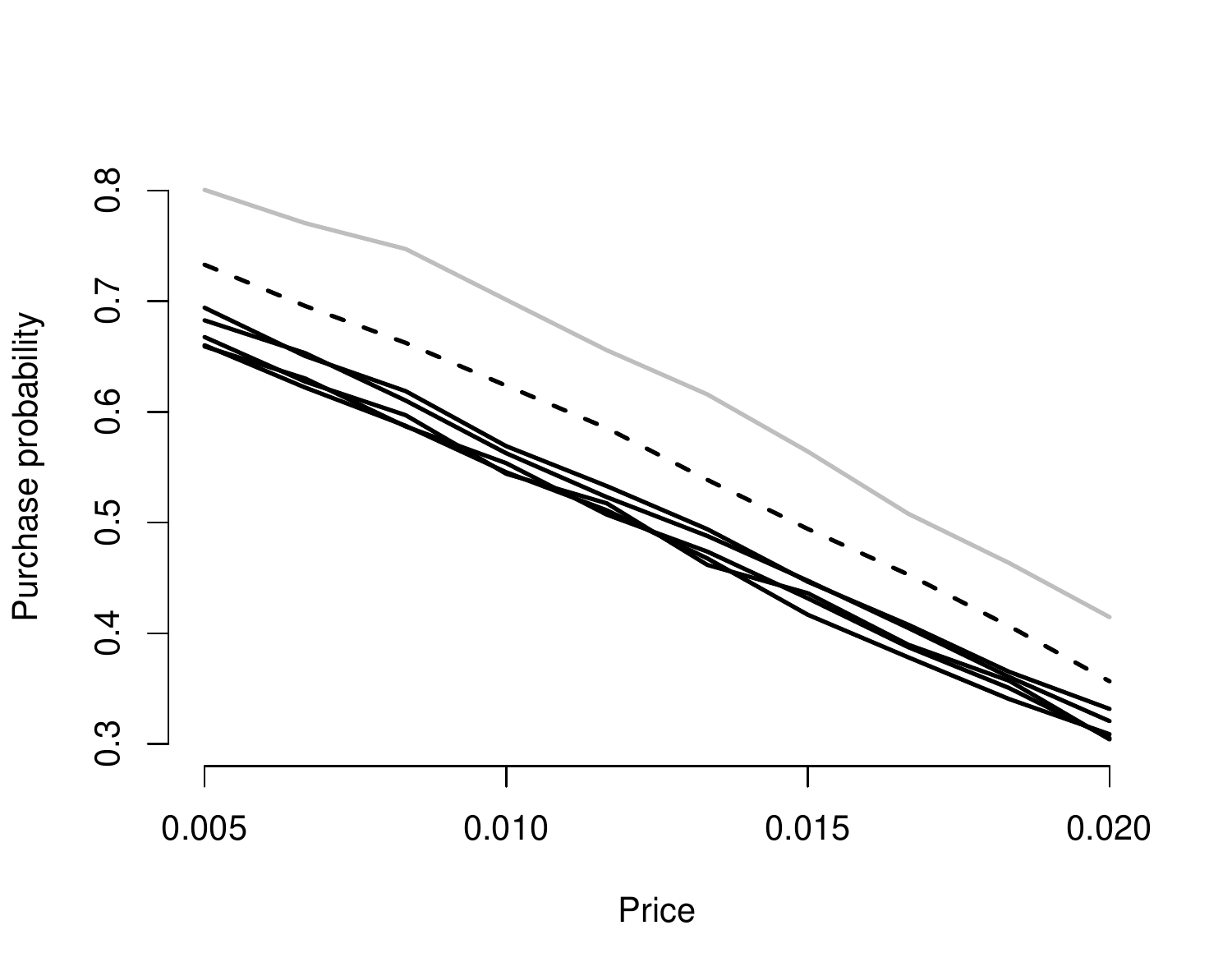}}\vspace{-7mm}
\singlespacing{\caption{Estimated purchase probabilities for {\sc All} brand
  detergent, with all other brands' prices fixed at the brand-specific
  mean observed price.
  The dashed curve uses {\sc All} as the base 
  category; the solid curves use all of the other possible 
  base categories.  The
  dotted curve results from an sMNP fit.  Although the model is fit as
a function of log-price, we display results as a function of dollars.}\label{fig:comp}}
\end{figure}

\cite{ivd, ivdComp} apply their methods to a consumer choice model of
clothing detergent 
purchases.  The data are available in their \texttt{MNP} package in
R.  We have records of purchasing decisions along with available
log-prices for shoppers choosing between {\sc All}, {\sc Era Plus},
{\sc Solo}, {\sc Surf},
{\sc Tide}, and {\sc Wisk} brand detergents.  There are 2657
observations and only six regression parameters, so we typically 
do not see large differences in estimated purchase
probabilities based on the various base category fits.  However,
specifying the base category to be 
{\sc All} --- which is rarely purchased despite its low price --- does give
somewhat different predictions for {\sc All} when its price is low.
We see this in Figure \ref{fig:comp}, where we set the prices for all
other brands at their 
brand-specific average, and consider predicted purchase probabilities
across a range of low prices for {\sc All}.   The predictions from
five of the base categories (solid curves) are very similar.  The
predictions when {\sc All} is the base category (dashed curve) are
notably higher.  When we apply the sMNP to the data, we see that its
predictions are intermediate to those of the various base category
fits (dotted curve).  In each case, the estimated purchase probability is routinely computed from the posterior predictive distribution.    

\begin{figure}[H] 
\centerline{\includegraphics[scale=0.45]{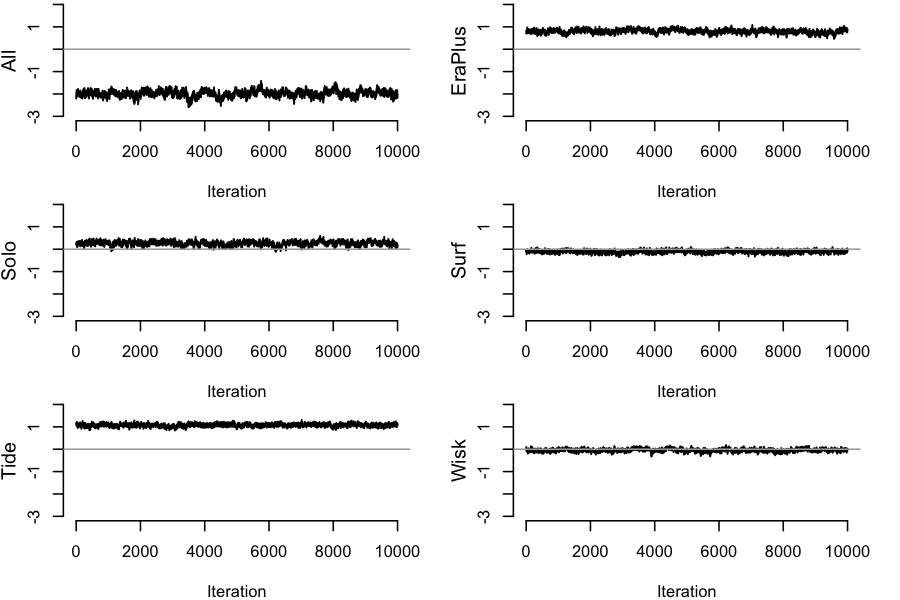}}\vspace{-5mm}
\singlespacing{\caption{Trace plots of samples from the posterior
  distributions of the intercept terms for an sMNP fit of the detergent data.}\label{fig:int}}
\end{figure}

\begin{figure}[H] 
\centerline{\includegraphics[scale=0.32]{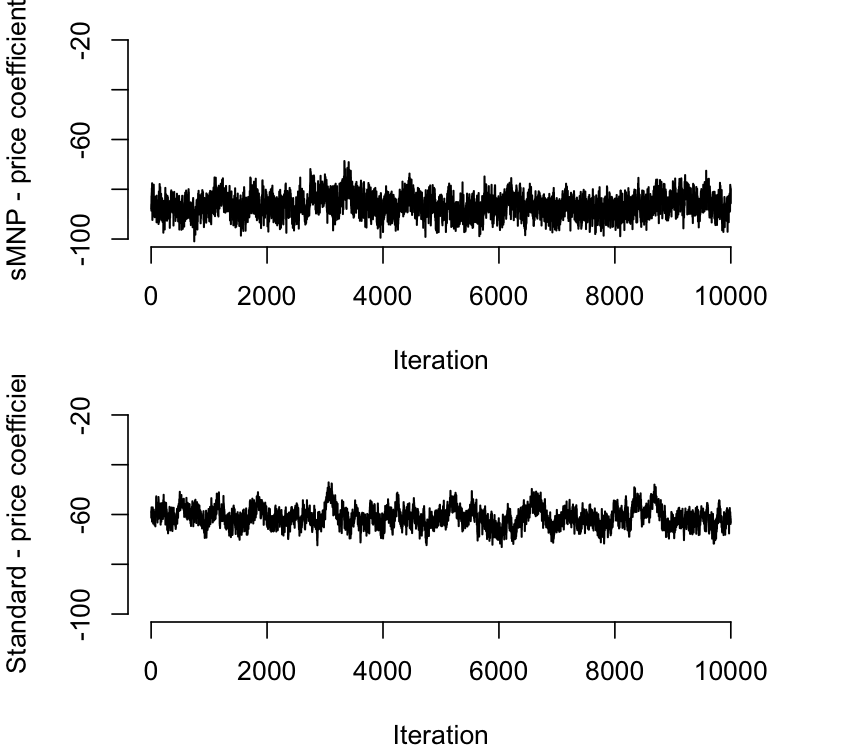}}\vspace{-5mm}
\singlespacing{\caption{Trace plots of samples from the posterior
  distributions of the price coefficients for sMNP and standard MNP 
  fits of the detergent data.}\label{fig:price}}
\end{figure}

To interpret the $\beta$ parameters, we know --- by 
the sum-to-zero property of
the intercept terms --- that a brand with an  intercept coefficient
that is persistently negative ({\sc All}) is less desirable than
average, in a sense (Figure \ref{fig:int}).  {\sc EraPlus} and {\sc Tide} 
are estimated to be more
desirable.  However, note that these intercepts do not reflect
marginal purchase probabilities, as less desirable brands may also
have lower prices.  
As economic theory would suggest, the price coefficient is
strongly negative (Figure \ref{fig:price}), which indicates that
raising a detergent's price 
(relative to the competitors) will lower its estimated purchase probability.    

Although these interpretations of the $\beta$ parameters are accurate, we
would argue that summaries of MNP results are best phrased in terms of
changes in posterior predicted selection probabilities.  For example,
one might consider the effect of a proposed price increase on the
current purchase probabilities.  We advocate this because predictions
take into account both $\beta$ and $\Sigma$ parameters, and the
$\Sigma$ parameters can be very difficult to interpret on their own.
If only the $\beta$ parameters are of interest in an application, we
would argue that a model that
assumes IIA may be more appropriate.  

\begin{figure}[H] 
\centerline{\includegraphics[scale=0.4]{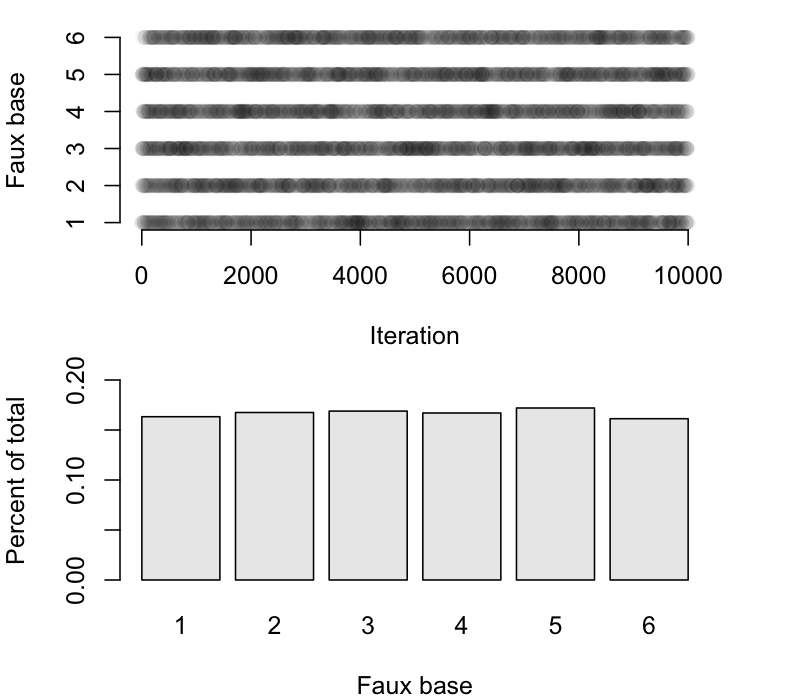}}\vspace{-5mm}
\singlespacing{\caption{Trace plot and histogram of samples from the posterior
  distribution of faux base parameters 
  $b$ for the detergent data.  In the upper panel,
  points are plotted with 2\% intensity.  The numbers 1 through 
  6 correspond to {\sc All}, {\sc EraPlus}, {\sc Solo}, {\sc Surf},
  {\sc Tide}, and {\sc Wisk}, respectively.}\label{fig:bPlot}}
\end{figure}

We also highlight the mixing behavior of the sMNP algorithm.
For example, the faux base parameter $b$ mixes extremely well,
as indicated by the near constant switching between its six possible
values (Figure \ref{fig:bPlot}). Further, the mixing of the price
parameter in the symmetric MNP algorithm compares favorably to 
the base category MNP in fits of these data (Figure \ref{fig:price}).
\cite{ivd} used these 
data to demonstrate improved mixing performance of their model 
relative to earlier MNP models, so these results are a comparison
against the state of the art.   

Here, the posterior of $b$ remains relatively flat. However, the extent that the data are informative about $b$ is precisely because the prior, for any fixed $b$, remains asymmetric. Consequently, in any finite data set, one of the base categories will look slightly ``better'' by the light of the prior predictive distribution {\em for some particular base category.} The fact that the data can inform us about $b$ is precisely why including $b$ in the model is necessary. Fixing $b$ at some arbitrary value, rather than moving the posterior probability of the ``faux bases'', would instead influence posterior inferences concerning the parameters of interest, such as choice probabilities themselves. Consequently, as a practical matter we do not recommend reporting posterior inferences on $b$, as they are a mere device for specifying a symmetric prior. \\


\subsection{Margarine purchases}

We also consider a similar analysis of consumer purchases of
margarine that are available in the \texttt{bayesm} package in R.
Again, our model only has intercepts and a price 
coefficient.  Following \cite{mcRossi}, we limit our analysis to
purchases of {\sc Parkay}, {\sc Blue Bonnet}, {\sc Fleischmann’s},
{\sc House} brand, {\sc Generic}, and {\sc Shedd Spread} tub
margarines.  And, following \cite{burgetteNord}, we limit the analysis
to the first purchase of one of these brands for each household.  This
results in a dataset with 507 observations.  With the smaller sample
size, there are larger differences in posterior estimated purchase
probabilities when one switches from one base category to another
in standard MNP fits.  

\begin{figure}[H] 
\centerline{\includegraphics[scale=0.7]{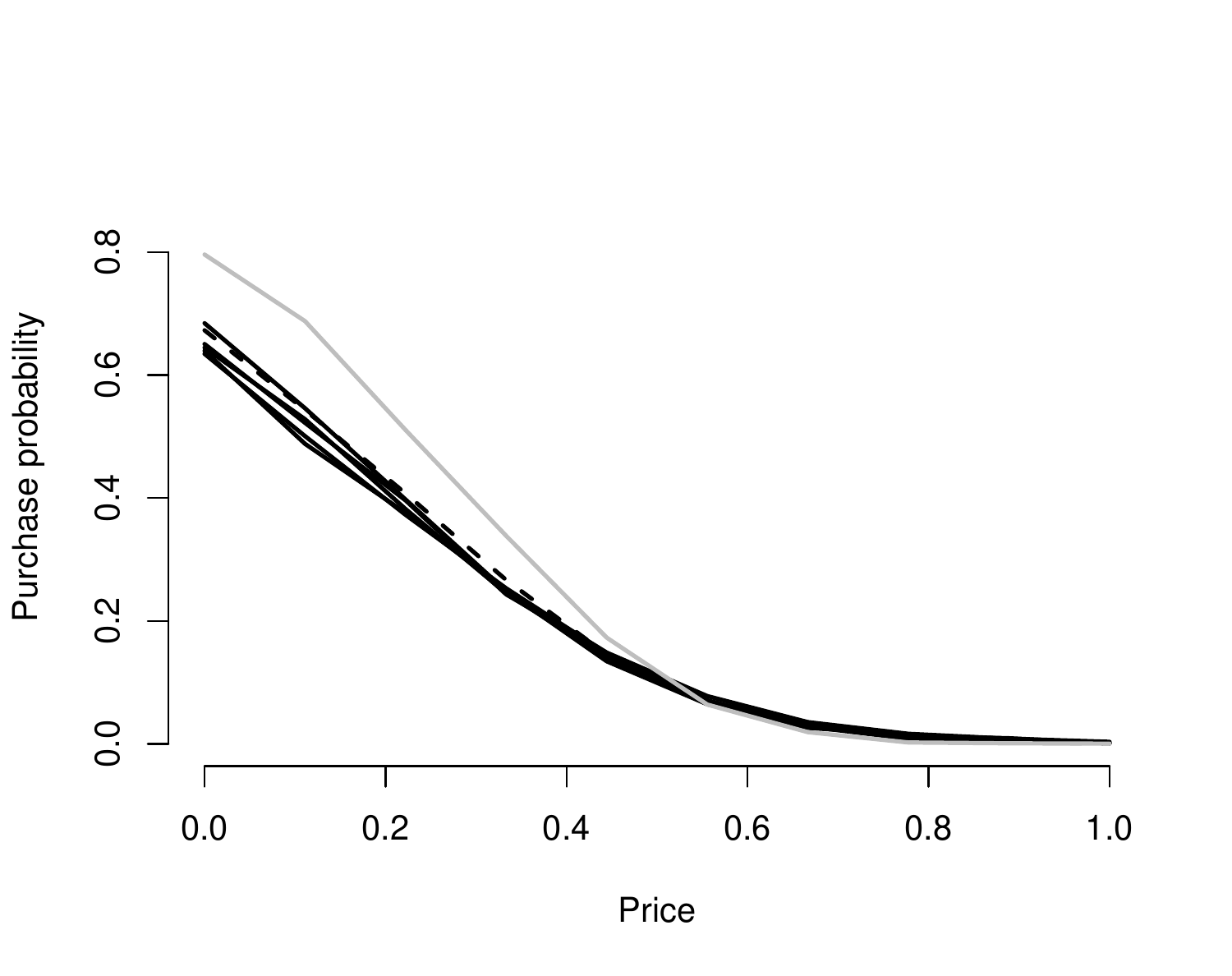}}\vspace{-7mm}
\singlespacing{\caption{Estimated purchase probabilities for {\sc House} brand
  margarine over a range of prices for that brand, with other prices
  fixed.  The solid curves are posterior predictions from standard MNP
  models, and the dashed curve is from the sMNP.  The gray curve uses
  {\sc House} brand as the base category.}\label{fig:housePriceProb}}
\end{figure}

In Figure \ref{fig:housePriceProb}, 
we see that sMNP predictions again tend to be between those of
standard MNP models when we consider all possible 
base categories, as was the case in 
Figure \ref{fig:comp}.  The observed {\sc House} brand prices are
between \$0.19 and \$0.64, so there is significant disagreement across
nearly the entire range of observed prices for that brand.  (With the
larger sample size in the detergent data, we only saw meaningful
differences when we extrapolated out of the observed price range.)  
Although
there is some Monte Carlo error in the estimates, it is insignificant
compared to the 19\% difference between the low and high estimates of
{\sc House}'s selection probability when it is priced at \$0.20.  

Thus, in both of these examples, we see that the sMNP gives
predictions that are between those of the standard MNP models that are
fit alternately with each base.  This is compatible with the heuristic
interpretation of the sMNP as a model that averages across base
categories in standard MNP models. 

An alternative approach to handling dependence on the base category
would be to fit an Imai and van Dyk-style MNP model using each base
category separately, and perform a post-hoc average of the fitted
probabilities.  We find this to be unappealing from several
perspectives.  First, the computation load is $p$ times as large as it
would be for a single, standard MNP fit; the sMNP is only slightly
more expensive than a single base 
category MNP.  More importantly, the sMNP constitutes a proper
Bayesian procedure, which automatically incorporates posterior
uncertainty about the base category and uses a
likelihood-weighted average of the possible models (bottom panel of
Figure \ref{fig:bPlot}).

%
%

\subsection{A simulation study}

Here we compare the fitted probabilities of MNP models that use each of the
possible base categories and the fitted probabilities that result from the
base category-free sMNP.  We simulate 50 datasets that are loosely
based on the consumer choice examples above.  We assume that $n=750$
consumers are choosing from $p =6$ products.  The simulated
product-specific intercepts and mean prices have correlation $0.9$ so
that more desirable products are more expensive, as one would expect.
The price coefficient was drawn uniformly from $[-1.25, -.75]$ so that
if a product is relatively less expensive, it will be more popular.
Finally, a 
$p \times p$ covariance matrix with expectation $I$ is drawn from an
inverse-Wishart distribution with 50 degrees of freedom.  The
simulation parameters were chosen so that each ``brand'' is chosen
with high probability.  Note that the data parameters were chosen
without regard to any set of identifying restrictions.  

\begin{figure}[H] 
\centerline{\includegraphics[scale=0.85]{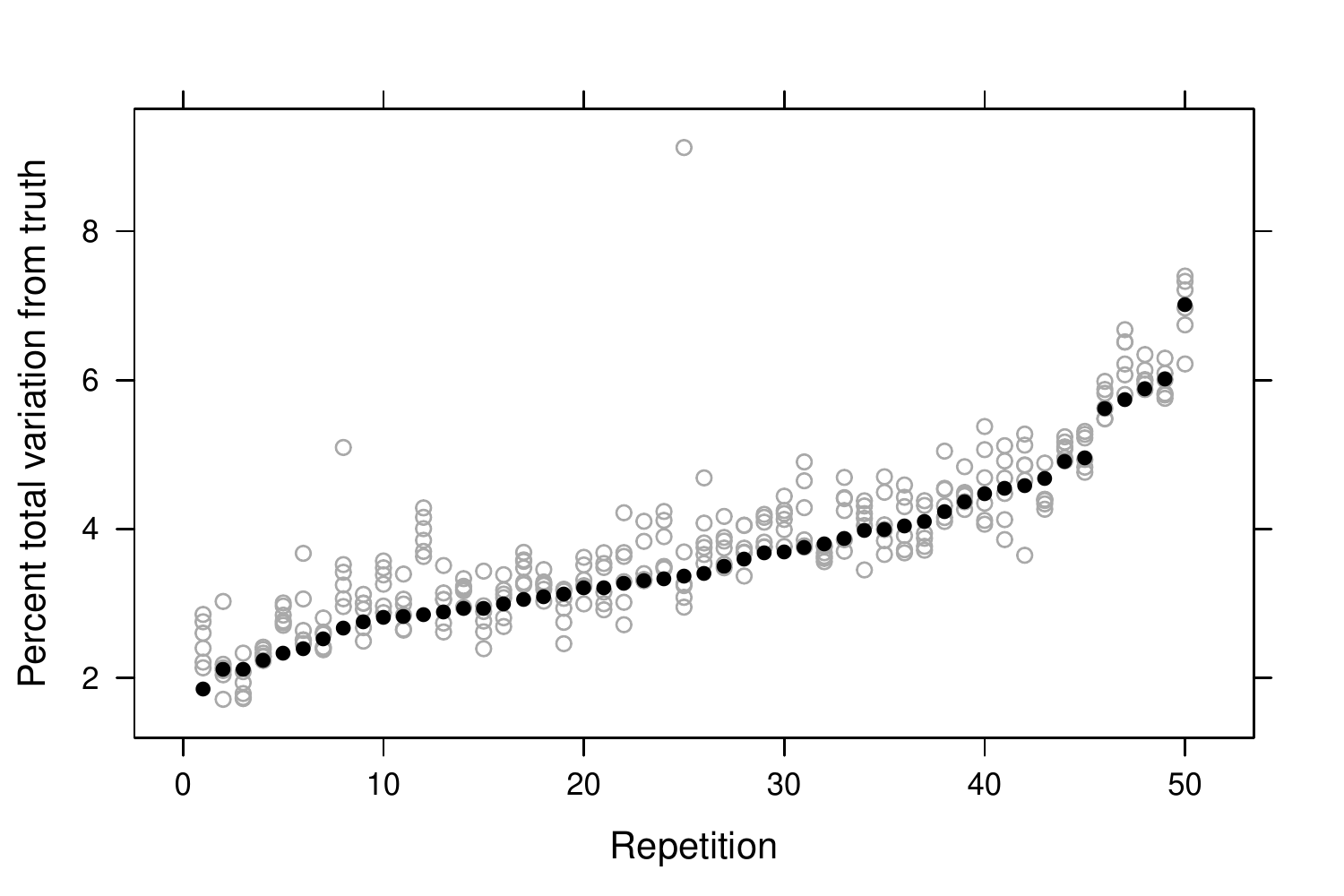}}\vspace{-7mm}
\singlespacing{\caption{Simulation results.  Points give the average percent total
  variation between true and estimated purchase probabilities.  Solid black
circles are from the sMNP.  Hollow gray circles are from MNP models that
use each of the six possible base category identifying restrictions.
The sMNP is almost never worse than every base category model, only 1 out of 50 cases, and in 41 out
of 50 cases it beats the median performing base category.}\label{fig:simRes}}
\end{figure}

We measure performance via the total variation between the estimated
and true purchase probabilities, averaged over the first 10 sets of
prices in each simulated dataset.  We expect that the sMNP will be
less prone to making ``extreme'' predictions in the sense of Figure
\ref{fig:housePriceProb}.  The results are summarized in Figure
\ref{fig:simRes}, and are consistent with this notion.  The plot
gives the average total variation from the 
true purchase probabilities for each of the base category MNP models
(hollow circles) and the sMNP (solid circles).  Note that the sMNP is
never the worst among the various base category models.  In 18 of
the 50 simulated cases, sMNP outperformed all of the base category
models.  In 41 out of 50 of the simulations, the sMNP
performed better than the median base category performance.  

\section{Identification}

A potential downside to our model is that it is not formally
identified.  In particular, the model would be identified if we were
able to restrict the trace of $\Sigma$, rather than the trace of
$\Sigma_{b}$.  If one of the diagonal elements of $\Sigma$ is
estimated to be substantially larger than 1, then the scale of $\beta$
will depend on $b$.  Although a fully identified model may be
preferable, we argue that little is lost in this case.   

First --- from the perspective of prior specification/elicitiation ---
the model is identified conditional on the discrete parameter $b$.  If
the analyst  wishes to specify an informative prior, this can be done
conditionally for each $b = 1,\ldots, p$.  If the model were only
identified conditional on a continuous working parameter, this process
becomes more difficult. Second --- on the side of interpretation ---
we would argue that $\beta$ parameters should be interpreted while
taking $\Sigma$ into account, and vice versa.  Since marginal
summaries do not do this, we feel that the best model summaries are
changes of fitted probabilities as a function of key outcome variables
such as in Figure \ref{fig:housePriceProb}, which are not impacted by
this identification 
issue.  If the analyst truly is interested in
features of the marginal posterior distribution of $\beta$ or
$\Sigma$, it is possible to post-process the results into a single,
identified scale by re-scaling the sampled values at each iteration of
the MCMC such that, for example, the trace of $\Sigma$ is equal to
$p$.  However, the signs of the estimated $\beta$ parameters are not
impacted by the under-identification of our model.  

Post-processing in order to identify Bayesian MNP models was popularized by
\cite{mcEtAl}, in the context of specifying a prior for $\tilde
\Sigma^*$, rather than the identified $\Sigma^*$.  As an aside, we
note that a related idea for solving the base category problem would
be to specify a full-rank inverse-Wishart prior for $\Sigma$,
without worrying about the conditional identifying restriction on the
location of the $W_i$.  However, this
approach proves to be numerically unusable.  
The $p$-dimensional inverse-Wishart
prior pushes the sampled values of $\Sigma$ toward the edge of the
parameter space, which quickly results in numerical problems that
result from sampling poorly-conditioned covariance matrices.

\section{Conclusion}

The analyses in this paper demonstrate
that careful handling of
the prior is necessary in order to obtain reliable predictions
from the Bayesian MNP.  As with any proper
Bayesian model, our estimates are biased, but they are not biased {\em
  against} any particular outcome category in the prior.  The same
can not be said of previous MNP models that estimate the covariance of
the utilities. 

With the prior
for the regression coefficients 
centered on zero, the sMNP estimates should be pulled toward
more moderate estimates.  Since multinomial data are quite coarse (in
the sense 
that each observation contributes little information compared to a
multivariate normal regression where the utilities are observed) 
we would argue that this prior-induced
regularization toward moderate predictions is highly desirable.  

When building more advanced  MNP models, symmetry may take on
even greater importance.  
For example, \cite{cripps} proposed an MNP
model that allows for a sparse representation of the precision matrix 
of the latent utilities.  However, they induce sparsity in the
precision of the base-subtracted utilities, not in the precision of the
original utilities.  This seems very likely to exacerbate the
problem of posterior estimates changing across different specifications
of the base category. Further, it is unclear that sparsity in the
base-subtracted precision corresponds to a 
meaningful data-generating process.  That said, it is likely that
favorable bias/variance tradeoffs can be made by specifying a
prior that pulls the precision toward a well-chosen, sparse
structure. 

More broadly, the regularizing effect of a Bayesian prior distribution
is at its most powerful when the likelihood is poorly behaved in some
way: when it is flat or spiky; when
identification is weak; when the number of parameters is large
relative to the sample size.  However, in each of these situations,
we should be worried that if our prior has undesirable features, they
may be preserved in the posterior.  
For example, MNP likelihoods can be quite flat, and therefore
the asymmetry of previously-proposed priors can propagate to the
posterior.   Data analysts may hope that such
undesirable features of the prior would be overwhelmed by the
likelihood.  This research suggests that while we cannot always count
on the data to cover flaws of our priors, we may be able to
design priors that lack the flaw in the first place, without giving up
computational tractability.

\section*{Appendix}

\subsection*{Full conditional distributions for the Gibbs sampler}

In the following sections, we provide expressions for the conditional distributions used in the Gibbs sampler.  The full conditionals that are standard distributions are determined by extracting relevant components from the joint distribution of all parameters.  The means and variances for the truncated univariate normal distributions follow from \cite{mcRossi}.

\subsubsection*{\underline{Step 1: Draw $\tilde{W} \mid Y, \tilde{\beta}_b, b, \Sigma_b, \alpha$}}

To sample $\tilde{W}$,
we iterate one-by-one through the elements of $\tilde{W}_{i,b}$. Note that $\tilde{w}_{i,b}$ is known given $b$ and $\tilde{W}_{i,b}$.
  After dropping the $b$th element of $\tilde{W}_i$ and the
  corresponding elements in $X_i$ and $\beta$, the full conditionals
  of elements of $\tilde{W}_{i,b}$ are truncated univariate normal.  The
  conditional means and variances can be calculated as described by
  \cite{mcRossi}, using $\tilde{\beta}_b$ as the coefficient vector and $\alpha^2\Sigma_b$ as the covariance. The truncations are: 
\begin{itemize}
\item If $Y_i = j\neq b$, sample $\tilde{w}_{ij}$ from a truncated
  normal so that $\tilde{w}_{ij} > -.5 \sum_{k \notin \{j, b\}} \tilde{w}_{ik}$ and
  $\tilde{w}_{ij} > \max(\tilde{w}_{ik} : k\notin \{j,b\})$.
\item If $Y_i \neq b$ and $Y_i = k \neq j$, sample $w^{\star}_{ij}$ from a
  truncated normal so that $\tilde{w}_{ij} < \tilde{w}_{ik}$ and $\tilde{w}_{ij} >
  -\sum_{l\neq b} \tilde{w}_{il} - \tilde{w}_{ik}$.
\item If $Y_i = b$, sample $\tilde{w}_{ij}$ from a truncated univariate
  normal such that $$\tilde{w}_{ij} < \min\{-.5\sum_{k\not\in \{b,j\}} \tilde{w}_{ik},
  -1(\max\{\tilde{W}_{-\{j,b\}}\} + \sum_{k\not\in \{b,j\}} \tilde{w}_{ik})\}.$$
\end{itemize}

\subsubsection*{\underline{Step 2: Draw $\tilde{\beta}_b \mid Y, b, \Sigma_b, \tilde{W}, \alpha$}}

The transformed coefficient vector is drawn according to a normal distribution.  This arises from the normal likelihood coupled with the normal prior specified for the coefficient vector, and standard Bayesian linear regression analysis specifies the conditional moments in closed form.  They are as follows:
\begin{eqnarray*}
\hat \beta_{b}& =& [\sum_{i=1}^n X_{i, b}^\top\Sigma_{b}^{-1}X_{i, b} + A^{-1}]
^{-1}   [\sum_{i=1}^n X_{i, b}^\top\Sigma_{b}^{-1}  \tilde W_{i,
  b}],  \\ 
\tilde \beta_{b} &\sim& \mbox{normal}\bigg(\hat \beta_b, \alpha^2\big (
\sum_{i=1}^n X_{i,  b}^\top\Sigma_{b}^{-1}X_{i, b} + A^{-1}\big ) ^{-1}\bigg).
\end{eqnarray*}
This update follows from the fact that if $\beta_b \sim \mbox{normal}(0, A)$, then $\tilde{\beta}_b \sim \mbox{normal}(0, \alpha^2 A)$.

\subsubsection*{\underline{Step 3: Draw $\alpha, \Sigma_b, b \mid Y, \tilde{\beta}_b, \tilde{W}$}}

The draw from this distribution is divided into a couple steps.  First, draw $b \mid \tilde \beta, \tilde W$.  Second, by working with the intermediate quantity $\tilde{\Sigma}_{b}$, draw $\tilde{\Sigma}_{b} \mid b, \tilde\beta, \tilde W$.   Once obtained, set $\alpha = \sqrt{\tr(\tilde\Sigma_{b})/(p-1)}$, followed by $\Sigma_b = \tilde{\Sigma}_{b}/\alpha^2$.  In this way, $\tr(\Sigma_b) = p - 1$ at each iteration as in \cite{burgetteNord}. 

Computationally, the major change from  \cite{burgetteNord} is the
draw from $(b, \tilde \Sigma_{b})$.  We derive this draw by first noting that from the full conditional we have: 
\begin{eqnarray*}
p(b, \tilde \Sigma_{b} | \mbox{all}) &\propto&
\exp\{-.5\sum (\tilde W_{i,b} - X_{i, b} 
\tilde \beta_{b})^\top \tilde \Sigma_{b}^{-1}(\tilde W_{i,b} - X_{i, b}
\tilde \beta_{b})\} \\
&\phantom{=} &\times |\tilde \Sigma_{b}|^{-n/2}  p(\tilde \Sigma_{b}|b) p(b)\\
& \propto &\exp\{-.5 \tr(\tilde
\Sigma^{-1}(S_{b} + \sum (\tilde W_{i,b} - X_{i, b}
\tilde \beta_{b})(\tilde W_{i,b} - X_{i, b}
\tilde \beta_{b})^\top))\} \\
&\phantom{=} &\times |\tilde \Sigma_{b}|^{-.5(n + \nu_0 +p)}.
\end{eqnarray*}Integrating over the variance, we obtain the multinomial draw for $b$ as:
\begin{eqnarray*}
p(b|\tilde \beta, \tilde W) &\propto &\int p(b, \tilde \Sigma_{b}
|\mbox{all}) d\tilde\Sigma_{b}\\ 
& \propto & |S_{b} + \sum (\tilde W_{i,b} - X_{i, b}
\tilde \beta_{b})(\tilde W_{i,b} - X_{i, b}
\tilde \beta_{b})^\top | ^{-(n + \nu_b)/2}
\end{eqnarray*}
Then, conditional on $b$, $\tilde \Sigma_{b}$ can be sampled from an
inverse-Wishart distribution as shown in the full conditional above and described by \cite{ivd}.

\bibliographystyle{ECA_jasa}
\bibliography{iMNPBib}

@article{jiao2015corrected,
  title={A corrected and more efficient suite of MCMC samplers for the multinomal probit model},
  author={Jiao, Xiyun and van Dyk, David A},
  journal={arXiv preprint arXiv:1504.07823},
  year={2015}
}

@article{albertChib,
  title={{Bayesian analysis of binary and polychotomous response data}},
  author={Albert, J.H. and Chib, S.},
  journal={Journal of the American Statistical Association},
  volume={88},
  number={422},
  pages={669--679},
  year={1993},
  publisher={American Statistical Association}
}

@article{burgetteNord,
title={{The trace restriction: An alternative identification strategy for the Bayesian multinomial probit model}},
author={Burgette, L.F. and Nordheim, E.V.},
journal={Journal of Business and Economic Statistics},
year = {2012},
pages={404--410},
volume = {30},
number = {3}
}

@book{rossiBook,
  title={{Bayesian Statistics and Marketing}},
year={2005},
  author={Rossi, PE and Allenby, GM and McCulloch, RE},
address={Chichester, West Sussex, England},
  publisher={Wiley}
}

@article{garrido,
  title={{Forecasting freight transportation demand with the space-time multinomial probit model}},
  author={Garrido, R.A. and Mahmassani, H.S.},
  journal={Transportation Research Part B: Methodological},
  volume={34},
  number={5},
  pages={403--418},
  year={2000},
  publisher={Elsevier}
}

@article{ivd,
  title={{A Bayesian analysis of the multinomial probit model using marginal data augmentation}},
  author={Imai, K. and van Dyk, D.A.},
  journal={Journal of Econometrics},
  volume={124},
  number={2},
  pages={311--334},
  year={2005},
  publisher={Elsevier}
}

@article{ivdComp,
  title={{MNP: R package for fitting the multinomial probit model}},
  author={Imai, K. and van Dyk, D.A.},
  journal={Journal of Statistical Software},
  volume={14},
  number={3},
  pages={1--32},
  year={2005}
}

@article{mcEtAl,
  title={{A Bayesian analysis of the multinomial probit model with fully identified parameters}},
  author={McCulloch, R.E. and Polson, N.G. and Rossi, P.E.},
  journal={Journal of Econometrics},
  volume={99},
  number={1},
  pages={173--193},
  year={2000},
  publisher={Elsevier}
}

@article{rudolph,
  title={{Who's responsible for the economy? The formation and consequences of responsibility attributions}},
  author={Rudolph, T.J.},
  journal={American Journal of Political Science},
  volume={47},
  number={4},
  pages={698--713},
  year={2003},
  publisher={Wiley Online Library}
}

@article{mcRossi,
  title={{An exact likelihood analysis of the multinomial probit model}},
  author={McCulloch, R. and Rossi, P.E.},
  journal={Journal of Econometrics},
  volume={64},
  number={1},
  pages={207--240},
  year={1994}
}

@article{tannerWong,
  title={{The calculation of posterior distributions by data augmentation}},
  author={Tanner, M.A. and Wong, W.H.},
  journal={Journal of the American Statistical Association},
  volume={82},
  number={398},
  pages={528--540},
  year={1987}
}

@book{train,
  title={{Discrete Choice Methods with Simulation}},
  author={Train, K.},
  year={2003},
  publisher={Cambridge University Press},
address={Cambridge}
}

@article{mcfadden74,
title={{The measurement of urban travel demand}},
author = {McFadden, D.},
journal = {Journal of Public Economics},
volume = {3},
number = {4},
pages = {303--328},
year = {1974}
}

@article{cripps,
  title={{Parsimonious estimation of the covariance matrix in multinomial probit models}},
  author={Cripps, E. and Fiebig, D.G. and Kohn, R.},
  journal={Econometric Reviews},
  volume={29},
  number={2},
  pages={146--157},
  year={2010},
  publisher={Taylor \& Francis}
}

@article{hoff,
  title={{Simulation of the Matrix Bingham--von Mises--Fisher distribution, with applications to multivariate and relational data}},
  author={Hoff, P.D.},
  journal={Journal of Computational and Graphical Statistics},
  volume={18},
  number={2},
  pages={438--456},
  year={2009},
  publisher={ASA}
}

\end{document}